\documentclass[aps,prl,twocolumn,superscriptaddress]{revtex4-1}

\usepackage{graphics}
\usepackage{graphicx}
\usepackage{color}
\usepackage{epstopdf}
\usepackage{amsmath}

\usepackage{ulem}

\begin{document}

\title{Liquid droplets act as ``compass needles'' for the stresses in a deformable membrane} 

\author{Rafael D. Schulman}
\affiliation{Department of Physics and Astronomy, McMaster University, 1280 Main St. W, Hamilton, ON, L8S 4M1, Canada.}
\author{Ren\'{e} Ledesma-Alonso}
\affiliation{CONACYT - Universidad de Quintana Roo, Boulevar Bah\'{i}a s/n, Chetumal, 77019 Quintana Roo, M\'{e}xico}
\affiliation{Laboratoire de Physico-Chimie Th\'eorique, UMR CNRS Gulliver 7083, ESPCI Paris, PSL Research University, 75005 Paris, France.}
\author{Thomas Salez}
\affiliation{Laboratoire de Physico-Chimie Th\'eorique, UMR CNRS Gulliver 7083, ESPCI Paris, PSL Research University, 75005 Paris, France.}
\affiliation{Global Station for Soft Matter, Global Institution for Collaborative Research and Education, Hokkaido University, Sapporo, Hokkaido 060-0808, Japan.}
\author{Elie Rapha\"{e}l}
\affiliation{Laboratoire de Physico-Chimie Th\'eorique, UMR CNRS Gulliver 7083, ESPCI Paris, PSL Research University, 75005 Paris, France.}
\author{Kari Dalnoki-Veress}
\email{dalnoki@mcmaster.ca}
\affiliation{Department of Physics and Astronomy, McMaster University, 1280 Main St. W, Hamilton, ON, L8S 4M1, Canada.}
\affiliation{Laboratoire de Physico-Chimie Th\'eorique, UMR CNRS Gulliver 7083, ESPCI Paris, PSL Research University, 75005 Paris, France.}

\date{\today}

\begin{abstract}
We examine the shape of droplets atop deformable thin elastomeric films prepared with an anisotropic tension. As the droplets generate a deformation in the taut film through capillary forces, they assume a shape that is elongated along the high tension direction. By measuring the contact line profile, the tension in the membrane can be completely determined. Minimal theoretical arguments lead to predictions for the droplet shape and membrane deformation that are in excellent agreement with the data. On the whole, the results demonstrate that droplets can be used as probes to map out the stress field in a membrane.

\end{abstract}


\maketitle
The physics of liquid droplets in contact with soft or deformable solids, elastocapillarity, is an active subject of research. Between capillary origami and wrinkling instabilities of thin films~\cite{Py2007, Patra2009,Van2010,Bae2015,Huang2007, Vella2010,King2012,Schroll2013,Andreotti2016};  the bending, coiling, and winding of slender structures~\cite{Roman2010,Rivetti2012,Fargette2014,Elettro2015,Sauret2017,Elettro2016,Schulman2017}; and elasticity-mediated propulsion of droplets~\cite{style2013patterning,karpitschka2016,liu2016}; there is no shortage of complexity, self-assembly, or beautiful examples of pattern formation in the field. In addition, some recent results have forced us to question familiar concepts of solid-liquid interactions.  For instance, studies on the partial wetting of liquid drops on soft solids show that Young's law is applicable on length scales much larger than the bulk elastocapillary length $\gamma/E$, where $\gamma$ is the liquid-air surface tension and $E$ is the Young's modulus of the solid. However, on smaller length scales, the contact line reveals a wetting ridge set by a Neumann construction involving surface stresses~\cite{Best2008,Jerison2011,Style2012,Style2013a,Marchand2012a,Park2014,Hui2014}. 

\begin{figure}[t!]
     \includegraphics[width=1.0\columnwidth]{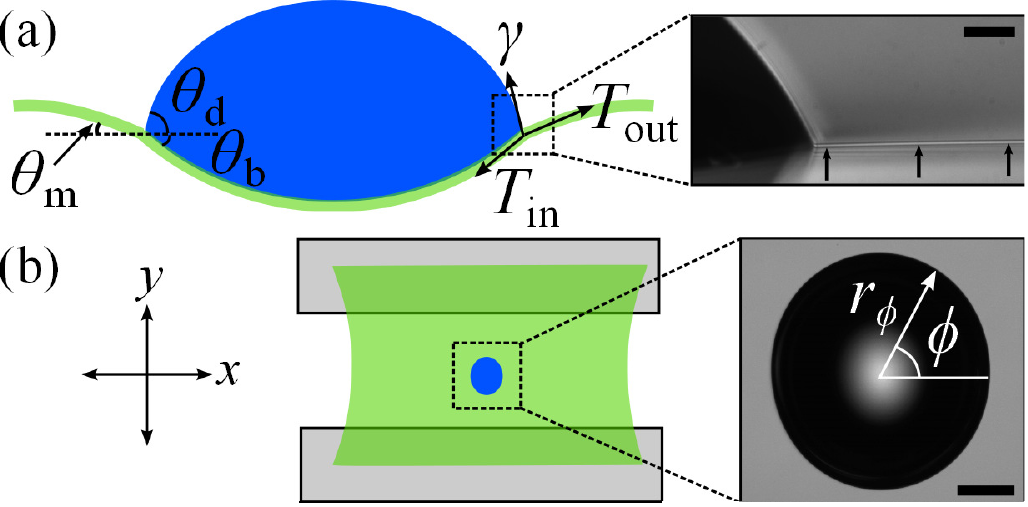}
\caption{(a) Schematic of the sideview of the drop/membrane system. The contact angle profile is determined by the force balance shown. On the right, an optical image of the contact line region taken along the $x$-direction is shown. The dark part is the liquid, the lighter gray part below is simply a reflection of the droplet off the film, and the thin light curve on the right (indicated by arrows) is the film itself. (b) Schematic of a free-standing elastomeric film between two supports with a single droplet atop. An optical topview image of the droplet is shown  on the right. Scale bars = 200 $\mu$m.}
\label{fig1}
\end{figure}

Partial wetting on deformable substrates may also be studied by  employing a highly compliant geometry, such as a droplet on a thin free-standing film~\cite{Shanahan85,Nadermann2013,Hui2015,Hui2015b,Schulman2016}. These studies have considered clamped films which are held taut and support a uniform and isotropic tension. As shown in Fig.~\ref{fig1}(a), the Laplace pressure of the droplet creates a bulge in the film below it, in the shape of a spherical cap, which is of the same order in size as the droplet itself. The deformations generated may be orders of magnitude larger than the bulk elastocapillary length, because stretching of the membrane is the relevant mode of elasticity ~\cite{Nadermann2013,Hui2015,Hui2015b,Schulman2016}. The contact line profile is determined by a Neumann construction, which incorporates both mechanical and interfacial tensions. This profile is characterized by the angles subtended by the liquid ($\theta_\mathrm{d}$) and bulge ($\theta_\mathrm{b}$) to the surrounding film (Fig.~\ref{fig1}(a)), which remains completely flat, i.e. the film's angle relative to the horizontal $\theta_m = 0$. From the Neumann construction, these angles are set by two parameters: the Young's angle $\theta_\mathrm{Y}$ of the same solid supported on a rigid substrate, and the ratio $T_\mathrm{in}/\gamma$, where $T_\mathrm{in}$ is the total mechanical and interfacial tension acting inside the contact region of the membrane/drop system~\cite{Schulman2016}. In the limit of infinite tension, the bulge vanishes and Young's law is recovered. 

In this study, we explore the partial wetting of a liquid droplet resting on an elastomeric membrane with an anisotropic tension. Surprisingly, the droplet assumes a shape which is elongated along the direction of high tension. We show from minimal theoretical considerations that the tensions in the film determine both the elongated shape of the wetting region and an observed  out-of-plane deformation of the film surrounding the droplet. Therefore, liquid droplets serve as non-destructive probes for mapping out the stress field in a membrane.

Free-standing elastomeric films of thickness $h \sim 240$~nm were prepared from Elastollan TPU 1185A (BASF) and suspended between two supports, as seen in Fig.~\ref{fig1}(b)~\cite{SI}. The sample is then stretched along the $y$-direction to produce a biaxial mechanical tension $T$, such that $T_y > T_x$. As can be seen in Fig.~\ref{fig1}(b), since the edges in the $x$-direction of the film are free, the membrane assumes a bowed configuration. However, at the center of the film, far from the edges, there is a biaxial tension which is uniform. In an experiment, a droplet of glycerol is placed near the center of the film. In doing so, we find that the droplet assumes a non-circular footprint which is elongated along the high tension direction (Fig.~\ref{fig1}(b)). Optical contact angle measurements are performed by viewing the sample from the side. These measurements are taken in two directions: viewing the sample from the $x$-direction, where the bulge and liquid-air interface can be simultaneously seen (as in Fig.~\ref{fig1}(a)); and viewing the sample from the $y$-direction. When observing from the $y$-direction, the supports block the view of the lower side of the film; hence, two separate droplets are needed, one on the top side and one on the bottom side of the film, to get images of both the bulge and liquid-air interface. Images of the two droplets are also taken from above, from which we obtain precise measurements of the contact radii $r_x$ and $r_y$, and hence the aspect ratio $\epsilon \equiv r_y/r_x$. We use droplets with contact radii in the range of 300 to 450 $\mu$m, small enough that gravity does not play a role but large enough that evaporation does not influence our measurements over the experimental time scale. The droplets are much smaller than the overall lateral size of the film.

\begin{figure}[t!]
     \includegraphics[width=1.0\columnwidth]{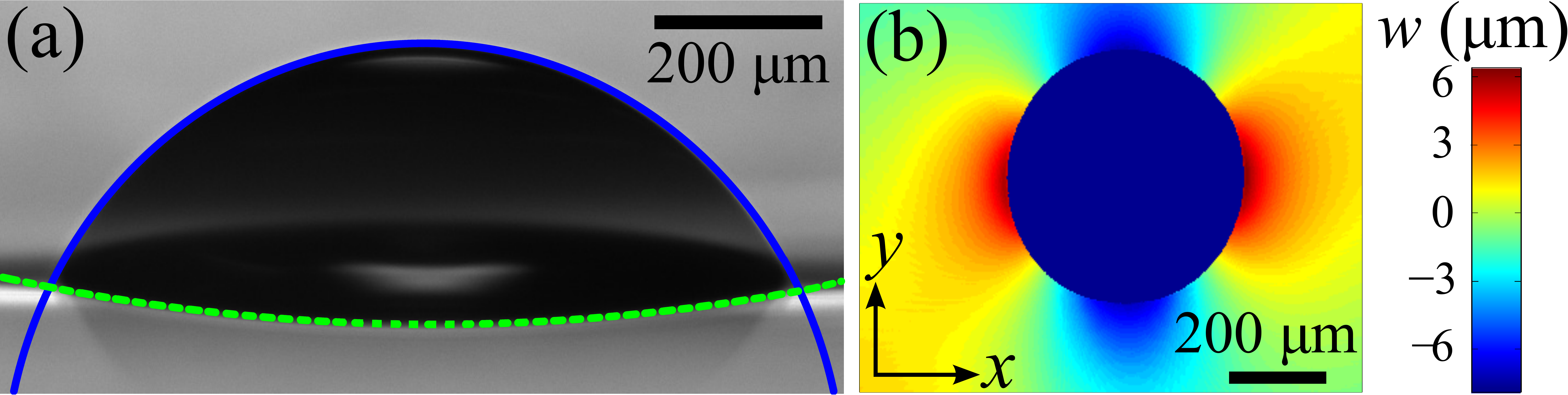}
\caption{(a) Optical sideview taken from the $x$-direction. The solid curve is a circular fit to the liquid-air interface and the dashed curve is a parabolic fit to the bulge. (b) Optical profilometry scan taken from above the droplet depicting the $z$-displacement $w$ of the film surrounding the wetting region (dark elliptical area). 
}
\label{fig2}
\end{figure}

A sample optical image taken from the $x$-direction is shown in Fig.~\ref{fig2}(a), where the droplet is sessile on top of the film, generating a bulge below it. The true 3D shape of the liquid-air interface is unknown. Nevertheless, its interface must have a constant mean curvature to ensure a constant internal pressure. However, to simplify our analysis, we fit all liquid-air interface profiles to circular caps and extract the average radii of curvature $R_{\mathrm{d},x}$ and $R_{\mathrm{d},y}$. As can be seen in by the solid curve in Fig.~\ref{fig2}(a), these fits exhibit excellent agreement with our experimental images~\cite{SI}. We note that $R_{\mathrm{d},y}>R_{\mathrm{d},x}$ in further support that the droplet's shape is not spherical. Knowing $r_i$ and $R_i$ in each direction $i = x,y$, the contact angles $\theta_{\mathrm{d},x}$ and $\theta_{\mathrm{d},y}$ (defined as the angles the liquid-air cap makes at the contact line relative to the $x$-$y$ plane as seen in Fig.~\ref{fig1}(a)) can be determined. Furthermore, the bulge profiles are fit to parabolas, the justification for which will be provided later.  As can be seen by the dashed curve in Fig.~\ref{fig2}(a), we find these fits to capture the bulge profile  well~\cite{SI}. From these fits, we determine $\theta_{\mathrm{b},x}$ and $\theta_{\mathrm{b},y}$, defined as the angles the bulge makes at the contact line relative to the $x$-$y$ plane, as defined in Fig.~\ref{fig1}(a).

We describe our films as membranes where bending can be ignored, and hence, the membrane slope is thought of as being discontinuous at the contact line. In reality, there exists a narrow region near the contact line where bending dominates  and the membrane curves to connect the bulge region to the outside region. However, since this bending region is too small to be measured in our experiments, the membrane description is appropriate. For droplets on membranes with isotropic tension, the film is flat outside the contact region, and the contact line shape is completely determined by $\theta_\mathrm{d}$ and $\theta_\mathrm{b}$~\cite{Nadermann2013,Hui2015,Hui2015b,Schulman2016}. However, for films with anisotropic tension, the film experiences an out-of-plane deformation outside the wetting region. Therefore, a complete picture of the contact line profile must include the angle of the membrane relative to the $x$-$y$ plane at the contact line which we denote with $\theta_\mathrm{m}$. As can be seen in Fig.~\ref{fig1}(a), the membrane curves down towards the droplet in the $y$-direction which we define to correspond to $\theta_\mathrm{m}>0$. Conversely, the film curves up towards the droplet in the $x$-direction and $\theta_\mathrm{m}<0$. Since $|\theta_\mathrm{m}|$ is small ($<4^\circ$) and the membrane is difficult to resolve optically, we employ optical profilometry (Veeco, Wyko NT1100) to probe the out-of-plane deformation of the membrane, $w$, where $w>0$ is defined to indicate the side from which the liquid droplet is placed. One such profilometry scan taken from the droplet side is shown in Fig.~\ref{fig2}(b). The membrane is pulled towards positive $w$ on the low tension side while being displaced towards negative $w$ on the high tension side.  From the profilometry data, it is straightforward to determine the values of $\theta_{\mathrm{m}}$ in the two principal directions: $x$ and $y$.  

Although the tension in the membrane is not known a priori, it may be determined using the contact line profile, as has been demonstrated in previous studies~\cite{Nadermann2013,Hui2015,Hui2015b,Schulman2016}. Through a Neumann construction, as depicted in Fig.~\ref{fig1}(a) where $T_\mathrm{in}$ and $T_\mathrm{out} = T_\mathrm{in}+\gamma\mathrm{cos}\theta_{\mathrm{Y}}$ contain mechanical and interfacial tensions (see Ref.~\cite{Schulman2016}), the contact line profile in a given direction is completely determined by the values of $T_\mathrm{in}/\gamma$ and $\theta_\mathrm{Y}$. The contact line profile is entirely characterized by the three internal angles subtended at the contact line: $\pi - \theta_\mathrm{d}-\theta_\mathrm{m}$, $\theta_\mathrm{d}+\theta_\mathrm{b}$, and $\pi - \theta_\mathrm{b}+\theta_\mathrm{m}$. The Neumann construction is only able to predict these internal angles; {\it not} $\theta_\mathrm{d}$, $\theta_\mathrm{b}$, and $\theta_\mathrm{m}$ individually. Therefore, using the value for $\theta_\mathrm{Y}$ and the angles $\theta_\mathrm{d}$, $\theta_\mathrm{b}$, and $\theta_\mathrm{m}$, we fit the internal angles to the Neumann construction prediction in $x$ and $y$ with $T_{\mathrm{in},x}/\gamma$ and $T_{\mathrm{in},y}/\gamma$ as the fitting parameters (see Ref.~\cite{SI}). The best fit allows us to extract values of $T_{\mathrm{in},x}/\gamma$ and $T_{\mathrm{in},y}/\gamma$.  A sample fit is provided in Table~\ref{tab1}, where it is clear that a single value of $T_{\mathrm{in},x}/\gamma$ captures the data well. The tensions measured in this way were found to be consistent with tensions computed from the strains in the film during stretching, attained using particle tracking in a separate experiment~\cite{SI}.

\begin{table}
\begin{ruledtabular}
\begin{tabular}{ c | c c }
 & Experiment & Best Fit \\  [1ex] \hline
$(\theta_{\mathrm{d},x}+\theta_{\mathrm{m},x})$ &  57.8 $\pm$ 1.2$^\circ$ & 59.5 $\pm$ 0.6$^\circ$ \\  
$(\theta_{\mathrm{d},x}+\theta_{\mathrm{b},x})$ & 76.6 $\pm$ 1.1$^\circ$ & 78.1 $\pm$  0.6$^\circ$ \\
$(\theta_{\mathrm{b},x}-\theta_{\mathrm{m},x})$ &  18.8 $\pm$ 0.8$^\circ$ & 18.7 $\pm$ 1.1$^\circ$  \\  
\end{tabular}
\end{ruledtabular}
\label{tab1}
\caption{\label{tab1} Sample fit of contact angle data in the $x$-direction to a Neumann construction to extract $T_{\mathrm{in},x}/\gamma$ = 2.7 $\pm$ 0.15.}
\end{table}

\begin{figure}[t!]
     \includegraphics[width=1.0\columnwidth]{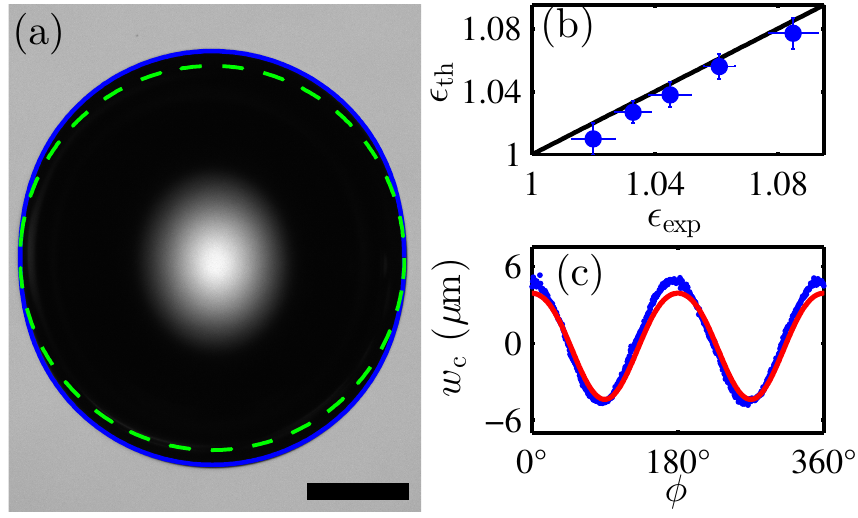}
\caption{(a) Topview image of an elongated droplet ($\epsilon \sim 1.09$) where the solid curve represents the best fit of Eq.~\ref{r_phi} and the dashed curve is a circle drawn for comparison. Scale bar = 200 $\mu$m. (b) A comparison between the experimental and theoretical values of $\epsilon$ for all samples. The line drawn represents $\epsilon_\mathrm{th} = \epsilon_\mathrm{exp}$. (c) Vertical position of the contact line around a sample droplet, where the points correspond to experimental data and the solid curve is calculated from Eq.~\ref{r_phi} and Eq.~\ref{w} with $A$ and $B$ obtained from bulge profile fits.
}
\label{fig3}
\end{figure}

We now turn to theoretical considerations, and begin with some simplifications to render the problem more tractable. First, we restrict our analysis to the case in which $\epsilon$ is close to unity, where the shape of the droplet may be treated as a perturbation to a spherical cap. As such, we make the assumption that the liquid-air interface profile as viewed along a general sightline oriented at $\phi$ (Fig.~\ref{fig1}(b)) is simply a circular cap with its own radius of curvature. This assumption is validated by the fact that such profiles are well fit to circular caps~\cite{SI}. Similarly, as will be justified later, all profiles of the bulge are parabolic in a good approximation. We make the simplistic assumption that the deformations produced by the two droplets on the film are only perturbative to the pre-tension. This assumption was found to be valid in a previous study done with isotropic tension~\cite{Schulman2016}, but is further supported by the fact that the contact angles are unchanged as additional droplets are placed onto the film as well as by the tension measurement done using particle tracking~\cite{SI}. Finally, we make small angle approximations when appropriate~\cite{footnote:small_angles}. To predict the shape of the droplet's footprint, i.e. projection of the wetted region onto the $x$-$y$ plane, we make use of the fact that the total height from the bottom of the bulge to the top of the droplet must be equal in every profile. We find that the contact radius  $r_\phi$ (defined in Fig.~\ref{fig1}(b)) of the footprint is given by (see Ref.~\cite{SI}):
%
%
 \begin{equation}
{r_\phi} \left[{\mathrm{arccos}  \Big( \mathrm{cos}\theta_\mathrm{Y}- \frac{\gamma}{2T_{\mathrm{in},\phi}}\mathrm{sin}^2\theta_\mathrm{Y}\Big) }\right]=C,
\label{r_phi}
\end{equation}
where $T_{\mathrm{in},\phi}$ is the total (mechanical and interfacial) tension in the $\phi$ direction in the region under the droplet, and $C$ is a constant which simply sets the overall length scale. We see that $r_\phi$ increases with $T_{\mathrm{in},\phi}$, consistent with the observation that droplets are elongated along the direction of highest tension. Since the membrane tensions are assumed to be unchanged by the addition of droplets, the tension remains purely biaxial with its principal axes aligned along $x$ and $y$, and $T_{\mathrm{in},\phi} = T_{\mathrm{in},x} \mathrm{cos}^2\phi + T_{\mathrm{in},y} \mathrm{sin}^2\phi $~\cite{Timoshenko1951}. In Fig.~\ref{fig3}(a), we show an optical topview of an elongated droplet on a film, where we also plot Eq.~\ref{r_phi} as a solid curve with $C$ = 514 $\mu$m found by fitting. We see that Eq.~\ref{r_phi} provides an excellent approximation of the elongated shape of the footprint. Furthermore, Eq.~\ref{r_phi} can be used to determine $\epsilon$ without any free-parameters. We see that the aspect ratio is high when $T_{\mathrm{in},y}$ is large while retaining a small $T_{\mathrm{in},x}$. For a quantitative comparison with our experimental observations, we refer to Fig.~\ref{fig3}(b), where all measurements of the aspect ratio $\epsilon_\mathrm{exp}$ are plotted against their predicted values  $\epsilon_\mathrm{th}$, computed using Eq.~\ref{r_phi} and the measured values of $T_{\mathrm{in},x}/\gamma$ and $T_{\mathrm{in},y}/\gamma$.  We find a good agreement between the experimental and theoretical values of $\epsilon$. We note that $\epsilon_\mathrm{th}$ is systematically smaller than $\epsilon_\mathrm{exp}$, which we attribute to the simplifications made in the theory.

\begin{figure*}[t!]
     \includegraphics[width=1.0\textwidth]{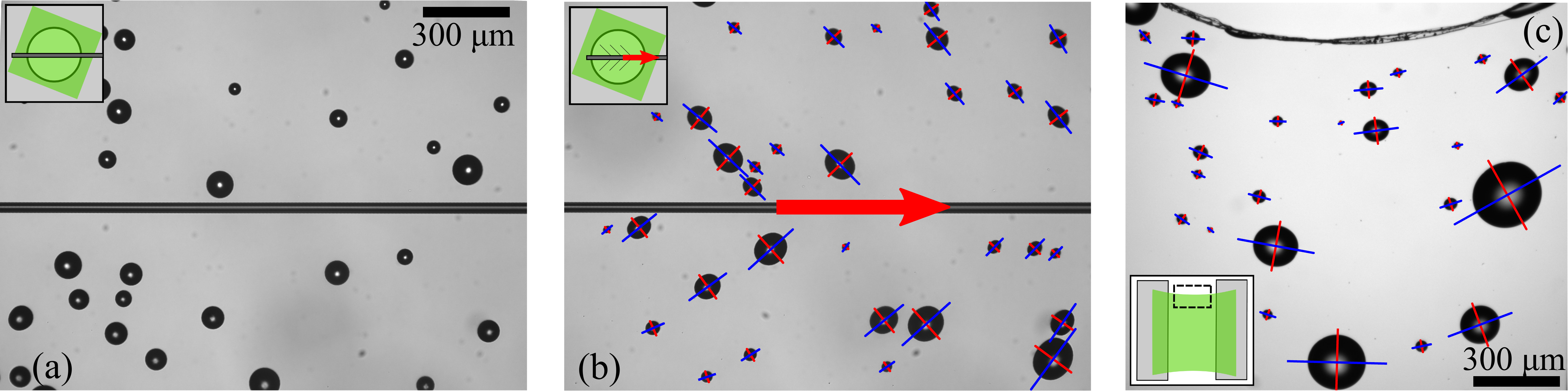}
\caption{(a) Droplets on a film with isotropic tension with a pipette laid across the film (schematized in the inset). (b) The pipette is moved in the direction indicated by the arrow to shear the film (schematized in the inset, where the thin diagonal lines indicate the principal direction of high tension).  (c) Droplets deposited near the boundary (see inset) of a film. The major/minor axes of the droplets are indicated by long/short lines in (b) and (c).
}
\label{fig4}
\end{figure*} 

To construct a full theoretical treatment of the membrane deformation, one may follow the approach laid out in articles by Davidovitch and co-workers, where the F\"{o}ppl-von K\'{a}rm\'{a}n equations are solved in the limit of negligible bending contributions~\cite{Davidovitch2011,King2012,Schroll2013}. However, further simplifications can be made when the deformation of the membrane by the droplet does not notably modify the pre-existing tension. Justified by our experimental observations, we have already made this assumption (in the claim that $T_{\mathrm{in},\phi}$ remains biaxial) to simplify the description of our system. As such, the deflection of a membrane ($w$) carrying a uniform biaxial tension of $T_{\mathrm{in},x}$, $T_{\mathrm{in},y}$  is given by~\cite{landau1986,SI}:
\begin{equation}
T_{\mathrm{in},x} \frac{\partial^2 w}{\partial x^2} + T_{\mathrm{in},y} \frac{\partial^2 w}{\partial y^2} = -  p(x,y),
\label{membrane_eq}
\end{equation}
where $p$ is the pressure distribution acting on the film. This equation is essentially Laplace's law but with anisotropic tension and in the limit of small membrane slopes. The small slope approximation is appropriate in this case since the bulge contact angles are always below 25$^\circ$. From this equation, it is straightforward to solve for the shape of the bulge, which forms in response to the uniform Laplace pressure over the wetted region.  We propose a solution of the form
\begin{equation}
w =Ax^2+By^2+w_0,
\label{w}
\end{equation}
where $w_0$ is an arbitrary vertical shift, and $A$ and $B$ are found experimentally~\cite{SI}. Thus, any vertical cross-section of the bulge passing through its apex is simply a parabola, which is the reason for fitting the bulge profiles with parabolas to extract $\theta_{\mathrm{b},x}$ and $\theta_{\mathrm{b},x}$. From these parabolic fits, the values of $A$ and $B$ are determined. Of course, we must still ensure Eq.~\ref{membrane_eq} is satisfied. Substituting $w$ back into Eq.~\ref{membrane_eq}, generates a criterion which can be written as~\cite{SI}:
\begin{equation}
\frac{T_{\mathrm{in},x}}{\gamma}\theta_{\mathrm{b},x}+\frac{T_{\mathrm{in},y}}{\gamma \epsilon}\theta_{\mathrm{b},y} = \mathrm{sin}\theta_{\mathrm{d},x}+\frac{1}{\epsilon}\mathrm{sin}\theta_{\mathrm{d},y} .
\label{consistency_eq}
\end{equation}
Evaluating both sides of Eq.~\ref{consistency_eq} using all our data yields a constant value of 1.75 $\pm$ 0.03 for the left side and 1.70 $\pm$ 0.06 for the right side, indicating that the data and theory are consistent.

A striking aspect of the system is the out-of-plane deformation of the membrane surrounding the wetting region, as evidenced in Fig.~\ref{fig2}(b). As shown for a sample droplet in Fig.~\ref{fig3}(c), the vertical position $w_\mathrm{c}$ of the contact line relative to its average value, oscillates with $\phi$. Knowing all parameters in Eqs.~\ref{r_phi} and~\ref{w}, we may attain a prediction for $w_\mathrm{c}$ by evaluating Eq.~\ref{w} at the position of the contact line given by Eq.~\ref{r_phi}. Therefore, we plot the prediction of the data in Fig.~\ref{fig3}(c) with a solid
curve, and find that it exhibits excellent agreement with the data.

As we have seen, droplets are elongated along the  high tension direction, i.e. the major and minor axes of the droplets' footprints align with the principal tension directions in the membrane~\cite{Timoshenko1951}. We can further test this property by a separate experiment where samples are made with different stress fields in which the principal directions are known. As shown in Fig.~\ref{fig4}(a), water droplets have been sprayed onto a free-standing elastomeric film on a circular washer, where the tension is completely isotropic and uniform. A thin glass pipette has then been placed into contact across the film, but does not significantly modify the film's stresses. As such, the droplets are completely round, in agreement with previous work done with isotropic tension~\cite{Schulman2016}. After these droplets have evaporated, we displace the pipette slightly towards the right as seen in Fig.~\ref{fig4}(b), which generates a shear stress in the membrane. If a membrane with isotropic stress is subjected to a shear stress $\tau_{xy}$, the principal directions of the stress are aligned at 45$^\circ$ to the $x$ and $y$ axes regardless of the shear's magnitude~\cite{Timoshenko1951}. To employ this known stress field as a test, we spray water droplets onto the film and fit the footprints of these with ellipses to extract the major/minor axes, which are displayed in Figs.~\ref{fig4}(b,c) as long/short lines. Computing the average angle that the major axes subtend to the pipette, we find 44 $\pm$ 8$^\circ$, as we would expect. Finally, at a free (i.e. stress-free) boundary, the principal directions are tangent (high tension) and normal (low tension) to the boundary~\cite{Timoshenko1951}. We thus purposely deposit liquid droplets near the bowed edges (see Fig.~\ref{fig1}(b)) of our stretched sample, as seen in Fig.~\ref{fig4}(c). Indeed, the major/minor axes plotted atop the droplets align with the principal directions.

We have performed experiments studying liquid droplets atop deformable membranes which carry an anisotropic tension. Droplets assume shapes which deviate from spherical caps and become elongated along the direction of highest tension. By measuring the contact line profile, we completely determine the tensions in the membrane. Using these tensions, along with a minimal theoretical model, we are able to form accurate predictions for the elongated shape of the droplet's footprint and the out-of-plane deformation of the membrane surrounding this region. Thus, liquid droplets may be used as a tool to map out the magnitudes and directions of the stresses in a membrane -- analogous to iron filings in magnetic fields.

 \begin{acknowledgements}
 The financial support by Natural Science and Engineering Research Council of Canada and the Global Station for Soft Matter, a project of Global Institution for Collaborative Research and Education at Hokkaido University, is gratefully acknowledged.
 \end{acknowledgements}


\begin{thebibliography}{36}
\bibitem [{foo()}]{footnote:small_angles}%
  \BibitemOpen
  \href@noop {} {}\bibinfo {note} {In the theoretical model, we make the
  approximation that $\mathrm{tan}\theta_\mathrm{b} \approx \theta_\mathrm{b}$
  and $\mathrm{tan}\theta_\mathrm{d}/2 \approx \theta_\mathrm{d}/2$. In our
  experiments, $\theta_\mathrm{b}<25^\circ$ and $\theta_\mathrm{d}/2<33^\circ$,
  so there is less than $\sim$10\% error in making these
  approximations.}\BibitemShut {Stop}%
\end{thebibliography}
\end{document}


\title{Supplemental Information for: ``Liquid droplets act as ``compass needles" for the stresses in a deformable membrane''} 

\author{Rafael D. Schulman}
\affiliation{Department of Physics and Astronomy, McMaster University, 1280 Main St. W, Hamilton, ON, L8S 4M1, Canada.}
\author{Ren\'{e} Ledesma-Alonso}
\affiliation{CONACYT - Universidad de Quintana Roo, Boulevar Bah\'{i}a s/n, Chetumal, 77019 Quintana Roo, M\'{e}xico}
\affiliation{Laboratoire de Physico-Chimie Th\'eorique, UMR CNRS Gulliver 7083, ESPCI Paris, PSL Research University, 75005 Paris, France.}
\author{Thomas Salez}
\affiliation{Laboratoire de Physico-Chimie Th\'eorique, UMR CNRS Gulliver 7083, ESPCI Paris, PSL Research University, 75005 Paris, France.}
\affiliation{Global Station for Soft Matter, Global Institution for Collaborative Research and Education, Hokkaido University, Sapporo, Hokkaido 060-0808, Japan.}
\author{Elie Rapha\"{e}l}
\affiliation{Laboratoire de Physico-Chimie Th\'eorique, UMR CNRS Gulliver 7083, ESPCI Paris, PSL Research University, 75005 Paris, France.}
\author{Kari Dalnoki-Veress}
\email{dalnoki@mcmaster.ca}
\affiliation{Department of Physics and Astronomy, McMaster University, 1280 Main St. W, Hamilton, ON, L8S 4M1, Canada.}
\affiliation{Laboratoire de Physico-Chimie Th\'eorique, UMR CNRS Gulliver 7083, ESPCI Paris, PSL Research University, 75005 Paris, France.}

\date{\today}

\begin{abstract}
\end{abstract}

\pacs{}

\renewcommand{\thefigure}{S\arabic{figure}}
\renewcommand{\theequation}{S\arabic{equation}}

\maketitle

\section{Sample Preparation}
Elastomeric films were prepared from Elastollan TPU 1185A (BASF). Solutions of Elastollan in cyclohexanone (Sigma-Aldrich) were prepared at 3\% weight fraction.  Upon spincoating these solutions, the Elastollan polymers, which contain hard and soft segments, assemble to form an elastomer with physical crosslinks. The Elastollan solutions were cast onto freshly cleaved mica substrates (Ted Pella Inc.) to produce highly uniform ($<$5\% variation) films with of thickness $h \sim $ 240~nm, measured using ellipsometry (Accurion, EP3). These films were subsequently heated at 100$^\circ$C for 90 min to remove any residual solvent from the elastomer. After annealing, these films were floated onto the surface of an ultrapure water bath (18.2 M$\Omega \cdot$cm, Pall, Cascada, LS) and picked up between two supports to form our sample. In our experiments, the liquid we use is glycerol (Caledon Laboratories Ltd.).

\section{Liquid Cap and Bulge Profiles}
\begin{figure}[h!]
     \includegraphics[width=17cm]{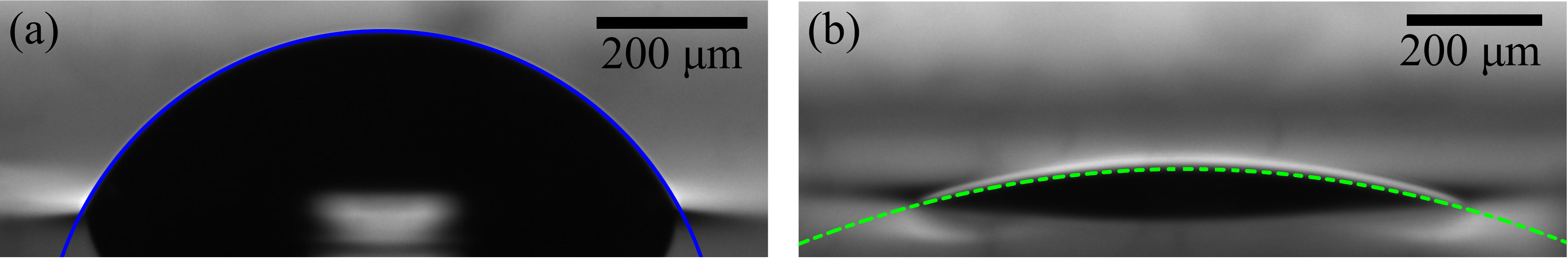}
\caption{(a) An optical sideview of a liquid-air interface profile taken from the $y$-direction, where the solid curve represents the circular cap fit to the profile. (b) An optical sideview of a bulge as seen from the $y$-direction. The dashed curve corresponds to the parabolic fit to the profile.}
\label{figS1}
\end{figure}

As explained in the main text, two droplets are placed on the film: one on each side. The purpose of doing so is to be able to visualize both the liquid-air interface and the bulge when viewing the sample from the $y$-direction, where the supports obscure our view of the lower side of the film. Sample optical microscopy images of of the two profiles when viewed from the $y$-direction are shown in Fig.~\ref{figS1}. In Fig.~\ref{figS1}(a), the solid curve represents the best fit of a circular cap to the liquid-air interface profile, and in Fig.~\ref{figS1}(b), the dashed curve represents the best fit of a parabola to the bulge profile. The fits desribe the optical profiles well. 

From optical profilometry, we find that the film is always pulled up towards the droplet on the low-tension side. This is seen in Fig.2(b) of the main text, where we observe a positive $w$ on the sides of the droplet along the $x$-direction. In fact, this small deformation can be visualized in Fig.~\ref{figS1}, where the film is seen to be pulled up towards the droplet in (a) and suppressed leading into the bulge in (b).

\begin{figure}
     \includegraphics[width=8.7cm]{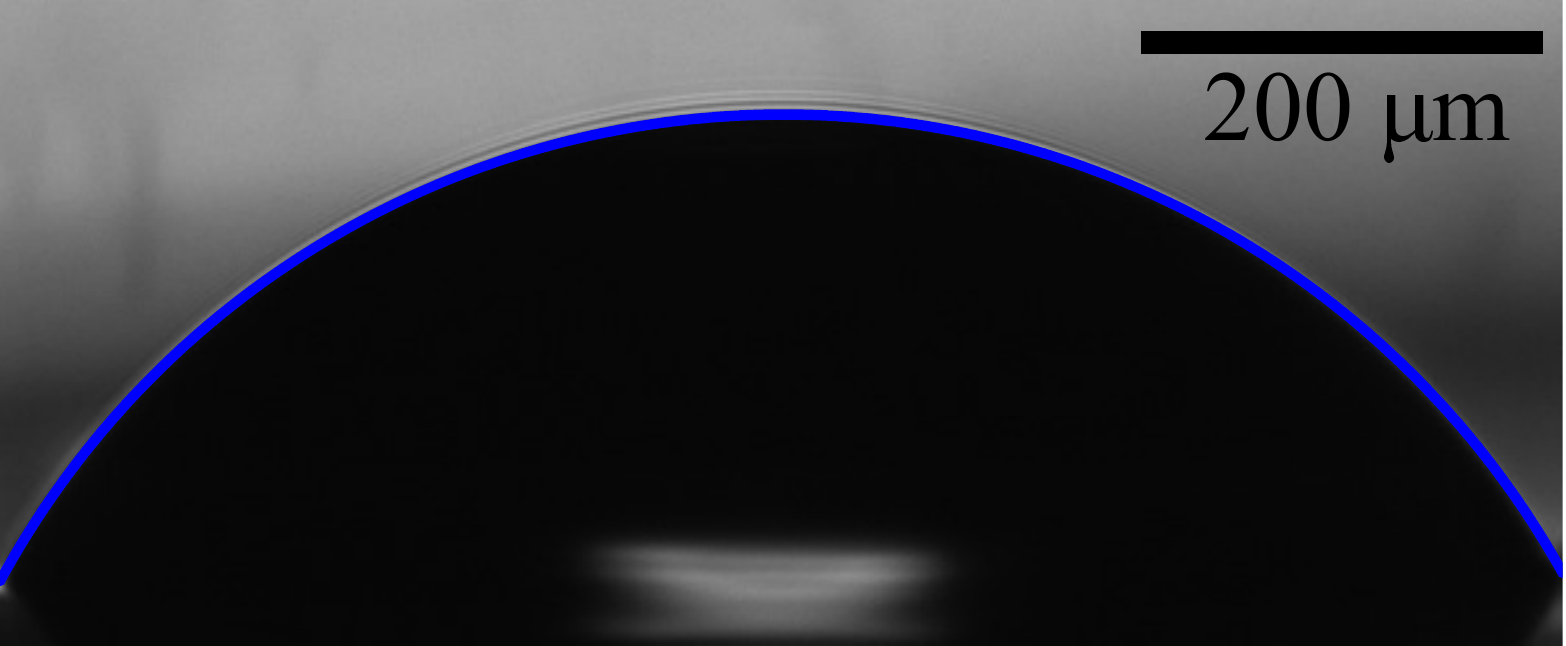}
\caption{An optical sideview of the liquid-air interface taken from a sightline oriented at $\phi = 45^\circ$ to the $x$-axis. The solid curve represents the circular fit to the profile.}
\label{figS2}
\end{figure}

As mentioned in the main text, to derive Eq.~1, we assume that the liquid-air interface profile is well described by a circular cap when viewed along any general direction oriented at $\phi$ to the $x$-axis. Experimentally, we observe that this assumption is fully reasonable. In Fig.~\ref{figS2}, we show a sample profile of a liquid-air interface taken along a sightline corresponding to $\phi = 45^\circ$, where we show the circular fit as a solid curve. As can be seen from this image, as well as any other sightline we have tested, the profile is always well described by a circular cap. Of course, since the 3D shape of the liquid-air interface is not spherical, the radius of curvature of the circular cap extracted from the profile fits changes along the various sightlines.

\section{Tension Verification}
Although the technique of using liquid contact angles to measure the tension in deformable membranes has been employed and verified in previous studies~\cite{Nadermann2013,Schulman2016}, we perform an additional validation here. In this experiment, we prepare a film where the tension is isotropic (droplets are completely round within experimental error) and is measured to be $T_{\mathrm{in}} /\gamma = 2.1 \pm 0.1$ using the contact angle technique from the main manuscript. Near the center of the film, we place small dirt particles to act as tracer particles. Next, we stretch the film along the $y$-axis, and from the tracer particles, the strains induced in the film in both directions are measured to be $e_x = -0.016 \pm 0.006$ and $e_y = 0.195 \pm 0.015$. Next, we perform contact angle measurements once again to determine the tension in this way. We find the change in tension from the initial state to be $\Delta T_{\mathrm{in},y} /\gamma  = 6.6 \pm 0.8$ and  $\Delta T_{\mathrm{in},x} /\gamma = 2.3 \pm 0.3$. If we suppose the interfacial tensions do not change appreciably upon stretching, the change in tension is purely mechanical.

To derive a theoretical expression for the change in tension upon straining the film, we employ Hooke's law~\cite{timoshenko1951}. We assume that there is no stress acting in the $z$-direction across the film, i.e. $\sigma_z = 0$. We also know that the mechnical tension is related to stress through film thickness $\Delta T = h \Delta \sigma$. Once again, the $\Delta$ signifies changes from the reference state of isotropic tension. As such, we may derive simple  expressions for the changes in mechanical tension upon stretching:

\begin{equation}
\Delta T_x = \frac{Eh(e_x+\nu e_y)}{1-\nu^2},
\label{delta_Tx}
\end{equation}
\begin{equation}
\Delta T_y = \frac{Eh(e_y+\nu e_x)}{1-\nu^2},
\label{delta_Ty}
\end{equation}
where $\nu$ is the Poisson ratio of the elastomer, which can be assumed to be 0.5 and $E$ is the Young's modulus. Some other quantities of interest to compute are the tension ratio 
\begin{equation}
(\Delta T_y/\Delta T_x) = \frac{e_y+\nu e_x}{e_x + \nu e_y},
\label{stress_ratio}
\end{equation}
since this quantity is independent of the modulus and film thickness, as well as the strain in the vertical direction, representing the fractional change in film thickness upon straining
\begin{equation}
e_z = \frac{\nu}{\nu-1}(e_x+e_y).
\label{z_strain}
\end{equation}

Substituting our measured strain values for the strains into Eq.~\ref{stress_ratio}, we find $(\Delta T_y/\Delta T_x) = 2.3 \pm 0.1$. This value is in agreement with the value measured using contact angles $(\Delta T_{\mathrm{in},y}/\Delta T_{\mathrm{in},x}) = 2.9 \pm 0.5$ within experimental error. In addition, as a consistency check, we verify Eq.~\ref{z_strain} by measuring the film thickness before and after stretching, to find a change in film thickness of -18.3\%, which agrees nicely with the predicted strain of -17.9\% from Eq.~\ref{z_strain}.

 To compare the individual values of the tension from contact angles against those from particle tracking, we must know $E$ and $\gamma$. We may find values for $E$ in the literature; nevertheless, this introduces error, as $E$ will depend on the details of the sample preparation for a physically cross-linked elastomer. With this caveat in place, we find measured literature values for the Young's modulus of Elastollan be roughly within the range $E = 9 \pm 1.5$ MPa~\cite{elastollan_modulus1,elastollan_modulus2,elastollan_modulus3}. The surface tension of glycerol is $\gamma = 0.063$ N/m~\cite{Lide2004}. For $h$ in Eqs.~\ref{delta_Tx} and~\ref{delta_Ty}, we use the stretched film thickness. As such, our predicted values from particle tracking are $\Delta T_x/\gamma = 3.1 \pm 0.9$ and  $\Delta T_y/\gamma = 7 \pm 2$, which compare well with the tensions determined from contact angles.

\section{Shape of the Wetting Region Perimeter}
In this section, we outline the arguments used to attain Eq.~1 in the main manuscript. We make some simplifying assumptions to render the problem more tractable. First and foremost, we assume that the shape of the liquid-air interface is a perturbation from a spherical cap. Given this assumption, there are two reasonable approximations which can be made. First, we approximate that any vertical cross-section of the liquid-air interface done along a general line oriented at $\phi$ (see Fig.~1(b) in the main manuscript) is a circular cap with its own radius of curvature. This assumption is found to be in agreement with the experimental observation that sideview profiles of the liquid-air interface from any such sightlines are well-described by circular caps (example seen in Fig.~\ref{figS2}). From the elliptical paraboloid shape of the bulge, which will be discussed in the following section, we know that all cross-sections of the bulge done in the same way are parabolas. Next, if we define $\phi'$ to be the angle subtended between the $x$-axis and the in-plane normal of the footprint perimeter, we make the approximation that $\phi' \approx \phi$. Of course, true equality only holds in the limit that the footprint shape is circular ($\epsilon = 1$), but it remains a good approximation for $\epsilon$ close to 1.

To begin the derivation, we point out that the vertical distance from the apex of the bulge to the top of the droplet, $h_\mathrm{tot}$, must be the same for any profile taken from a cross-section along a line oriented at $\phi$. For a given $\phi$, this height is found as the sum of the liquid cap height $h_{\mathrm{d},\phi}$ and the bulge height $h_{\mathrm{b},\phi}$. The height of the liquid cap can be found through a simple circular cap identity $h_{\mathrm{d},\phi} = r_\phi \mathrm{tan}\theta_{\mathrm{d},\phi}/2$, where $r_\phi$ is the footprint radius and $\theta_{\mathrm{d},\phi}$ is the cap's contact angle, both for this value of $\phi$. A similar relationship can be found for the parabola which represents the bulge, $h_{\mathrm{b},\phi} =\frac{ r_\phi}{2} \mathrm{tan}\theta_{\mathrm{b},\phi}$. Therefore, we can write

\begin{equation}
h_{\mathrm{tot}} = r_\phi \Big(\frac{1}{2}\mathrm{tan}\theta_{\mathrm{b},\phi}+\mathrm{tan}\frac{\theta_{\mathrm{d},\phi}}{2}\Big).
\end{equation}
Our final simplification in this derivation is to assume that $\theta_\mathrm{b}$ and $\theta_\mathrm{d}/2$ are small angles, such that $\mathrm{tan}\theta_\mathrm{b} \approx \theta_\mathrm{b}$ and $\mathrm{tan}\theta_\mathrm{d}/2 \approx \theta_\mathrm{d}/2$. This approximation is reasonable for our experiments where $\theta_{\mathrm{b},\phi} < 25^\circ$ and $\theta_{\mathrm{d},\phi}/2 < 33^\circ$, so there is less than 11\% error in making this approximation at this point. Generally these assumptions become increasingly more appropriate the smaller $\theta_\mathrm{Y}$ is. Employing the small-angle limit, we are left with
\begin{equation}
h_{\mathrm{tot}} = \frac{r_\phi}{2} \Big(\theta_{\mathrm{b},\phi}+\theta_{\mathrm{d},\phi}\Big).
\label{h_tot1}
\end{equation}
The angle sum $\theta_{\mathrm{b},\phi}+\theta_{\mathrm{d},\phi}$ represents the internal angle of the liquid at the contact line, and can be simply predicted using a Neumann construction, as was shown in previous work~\cite{Schulman2016}.
\begin{equation}
\theta_{\mathrm{b},\phi}+\theta_{\mathrm{d},\phi} = {\mathrm{arccos}  \Big( \mathrm{cos}\theta_\mathrm{Y} - \frac{\gamma}{2T_{\mathrm{in},\phi}}\mathrm{sin}^2\theta_\mathrm{Y}\Big) }.
\end{equation}
This Neumann construction should be set up normal to the contact line. However, since $\phi' \neq \phi$ as outlined before, the normal line does not pass through the droplet center. Therefore, to simplify the problem, we assume that $\phi' \approx \phi$, which implies that the Neumann construction above represents the internal angle of the liquid for a cross-section taken from the contact line to the droplet center, oriented at an angle $\phi$ to the $x$-axis. Thus, noting that $h_\mathrm{tot}$ in Eq.~\ref{h_tot1} is just some constant value, we arrive at the final result
\begin{equation}
r_\phi = \frac{C}{\mathrm{arccos}  \Big( \mathrm{cos}\theta_\mathrm{Y} - \frac{\gamma}{2T_{\mathrm{in},\phi}}\mathrm{sin}^2\theta_\mathrm{Y}\Big) },
\label{r_phi}
\end{equation}
where $T_{\mathrm{in},\phi}$ is the total tension in the $\phi$ direction in the region under the droplet, and $C$ is a constant which simply sets the overall scale of the region. As discussed in the main text, we assume that the deformations
produced by the two liquid drops on the film are only perturbative to the pre-tension of the membrane. This assumption was validated in  previous study done with isotropic tension~\cite{Schulman2016}, but is further supported
by the experimental observation that the contact angles remain constant as additional
droplets are placed onto the film and also by the tension confirmation using tracer particles described in the previous section. Since we have prepared the film to have a biaxial tension with principal directions in $x$ and $y$, the tension in any direction $\phi$ is $T_{\mathrm{in},\phi} =  T_{\mathrm{in},x} \mathrm{cos}^2\phi + T_{\mathrm{in},y} \mathrm{sin}^2\phi$~\cite{timoshenko1951}.

\section{Derivation of the Bulge Shape}
The small-slope out-of-plane deformation of a film $w\left(x,y\right)$ subjected to a transverse pressure is described by the F\"{o}ppl-von K\'{a}rm\'{a}n equations~\cite{landau1986}:

\begin{subequations}
\begin{equation}
D\Delta^2 w - h \nabla \cdot \big( \sigma \cdot \nabla w \big)\ = p , \\
\label{fvk1}
\end{equation}
\begin{equation}
\nabla \cdot \sigma = 0 \ , 
\label{fvk2}
\end{equation}
\end{subequations}
where $p$ is the transverse pressure distribution, $D$ is the flexural rigidity, $h$ is the film thickness, and $\sigma$ is the stress tensor. In our system, bending can be neglected, and the first term in Eq.~\ref{fvk1} can be ignored. Motivated by experimental observations, we make the simplifying assumption that the deformation of the membrane by the droplet does not notably modify the pre-existing tension. Therefore, the stress in the membrane is the as-prepared biaxial stress which is uniform in the region near the center of the film where the experiment is performed. Since the stress is uniform, Eq.~\ref{fvk2} is automatically satisfied, and Eq.~\ref{fvk1} can be simplified to:
\begin{equation}
T_{\mathrm{in},x}\dfrac{\partial^2 w}{\partial x^2} + 2T_{\mathrm{in},xy}\dfrac{\partial^2 w}{\partial x \partial y}   +T_{\mathrm{in},y}\dfrac{\partial^2 w}{\partial y^2}= - p , \\
\end{equation}
where we have used the general relation that tension is stress multiplied by film thickness. Since the film is prepared with a biaxial tension in which the principal directions are aligned with $x$ and $y$, it implies that $T_{\mathrm{in},xy} = 0$, thus we are left with
\begin{equation}
T_{\mathrm{in},x}\dfrac{\partial^2 w}{\partial x^2}+T_{\mathrm{in},y}\dfrac{\partial^2 w}{\partial y^2}=- p\left(x,y\right) \ ,
\label{Eq:Membrane}
\end{equation}
Since there is no flow within the liquid, the droplet must contain a uniform pressure, so $p$ is a constant value in our case and given by the Laplace pressure of the droplet $p =- \gamma \big(\frac{1}{R_{\mathrm{d},x}}+\frac{1}{R_{\mathrm{d},y}} \big)$, where the negative sign indicates that the pressure acts on the film in the negative $z$-direction.
Note that Eq.~\ref{Eq:Membrane} is simply the anisotropic Laplace's law in the limit of small slopes.

For the particular case in which $T_{\mathrm{in},x}=T_{\mathrm{in},y}$, we recover the isotropic Laplace's law (small slopes), for which the solution is given by
\begin{equation}
w\left(r\right)=\dfrac{pr^2}{4T_{\mathrm{in},x}}+c_0\ln(r)+c_1 \ ,
\end{equation}
with $r=\sqrt{x^2+y^2}$.
Since the position of the membrane at $x=y=0$ must be finite, we find that $c_0=0$, whereas $c_1=w_0$, an arbitrary vertical shift of the system.
Therefore, the isotropic solution becomes
\begin{equation}
w\left(x,y\right)=\dfrac{p}{4T_{\mathrm{in},x}}\left(x^2+y^2\right)+w_0 \ ,
\end{equation}

For the anisotropic case, since $\epsilon$ is near 1, we expect that the solution should be very similar to the isotropic case. Thus, we propose
\begin{equation}
w\left(x,y\right)=Ax^2+By^2+Cxy+Dx+Ey+F \ .
\end{equation}
In addition, we know that $\partial_x w=\partial_y w=0$ at $x=y=0$, which provides the values $D=0$ and $E=0$.
Once more, at $x=y=0$ an arbitrary vertical shift of the system $w_0$ is asigned to the coefficient $F$.
Additionally, we must consider the symmetry conditions: 1) $\partial_x w=0$ at $x=0$; 2) $\partial_y w=0$ at $y=0$; both implying that $C=0$.

Therefore, we have
\begin{equation}
w\left(x,y\right)=Ax^2+By^2+w_0 \ .
\end{equation}
Plugging this ansatz back into Eq.~\ref{Eq:Membrane} leads to
\begin{equation}
2T_{\mathrm{in},x}A + 2T_{\mathrm{in},y}B = \gamma \Big(\frac{1}{R_{\mathrm{d},x}}+\frac{1}{R_{\mathrm{d},y}} \Big)  \ .
\label{laplace1}
\end{equation}
We can also write $A$ and $B$ in terms of $\theta_{\mathrm{b},x}$ and $\theta_{\mathrm{b},y}$. At $x = r_{x}$, $y = 0$ we have $\partial_x w = 2Ar_x = \mathrm{tan}\theta_{\mathrm{b},x} \approx \theta_{\mathrm{b},x}$. Similarly, $2Br_y \approx \theta_{\mathrm{b},y}$.  In addition, a circular cap identity can be used to write $R_{\mathrm{d},x}$ and $R_{\mathrm{d},y}$ in terms of $\theta_{\mathrm{d},x}$ and $\theta_{\mathrm{d},y}$. Thus, Eq~\ref{laplace1} becomes
\begin{equation}
\frac{T_{\mathrm{in},x}\theta_{\mathrm{b},x}}{r_x} + \frac{T_{\mathrm{in},y}\theta_{\mathrm{b},y}}{r_y} = \gamma \Big(\frac{\mathrm{sin}\theta_{\mathrm{d},x}}{r_x}+\frac{\mathrm{sin}\theta_{\mathrm{d},y}}{r_y} \Big)  \ .
\label{laplace2}
\end{equation}
Finally, we arrive at Eq.~4 in the main manuscript 
\begin{equation}
\frac{T_{\mathrm{in},x}\theta_{\mathrm{b},x}}{\gamma} + \frac{T_{\mathrm{in},y}\theta_{\mathrm{b},y}}{\gamma \epsilon} =  \mathrm{sin}\theta_{\mathrm{d},x}+\frac{1}{\epsilon}\mathrm{sin}\theta_{\mathrm{d},y}   \ .
\label{laplace3}
\end{equation}

\section{Neumann construction}
The three internal angles characterizing the contact line profile are $\pi - \theta_\mathrm{d}-\theta_\mathrm{m}$, $\theta_\mathrm{d}+\theta_\mathrm{b}$, and $\pi - \theta_\mathrm{b}+\theta_\mathrm{m}$. Thus, the internal angles are set by three different linear combinations of the measured angles: $\theta_\mathrm{d}+\theta_\mathrm{m}$, $\theta_\mathrm{d}+\theta_\mathrm{b}$, and $\theta_\mathrm{b}-\theta_\mathrm{m}$. As was done in Ref.~\cite{Schulman2016}, one may apply a Neumann construction to attain predictions for these three angle combinations:

\begin{subequations}
\begin{align}
\cos\left(\theta_d+\theta_m\right) &=\dfrac{\left[T_{out}/\gamma\right]^2-\left[T_{in}/\gamma\right]^2+1}{2\left[T_{out}/\gamma\right]}  , \\
\cos\left(\theta_b+\theta_d\right) &=\dfrac{\left[T_{out}/\gamma\right]^2-\left[T_{in}/\gamma\right]^2-1}{2\left[T_{in}/\gamma\right]} , \\
\cos\left(\theta_b-\theta_m\right) &=\dfrac{\left[T_{out}/\gamma\right]^2+\left[T_{in}/\gamma\right]^2-1}{2\left[T_{out}/\gamma\right]\left[T_{in}/\gamma\right]} ,
\end{align}
\label{Eqs:Cos}
\end{subequations} 
where $T_\mathrm{out} = T_\mathrm{in}+\gamma\mathrm{cos}\theta_\mathrm{Y}$. These predictions depend only on two parameters: $T_{\mathrm{in}}/\gamma$ and $\theta_\mathrm{Y}$. Of course, this Neumann construction may be carried out normal to the contact line at any point along the perimeter.  Thus, using the value we measure for $\theta_\mathrm{Y}$, we fit these predictions separately to the measured internal angles in $x$ and in $y$ to find the best fit parameter values of $T_{\mathrm{in},x}/\gamma$ and $T_{\mathrm{in},y}/\gamma$. The best fitted values of $\theta_\mathrm{d}+\theta_\mathrm{m}$, $\theta_\mathrm{d}+\theta_\mathrm{b}$, and $\theta_\mathrm{b}-\theta_\mathrm{m}$ are listed in Table~I of the main manuscript. 

\bibliography{Schulman2016bib}